\documentstyle[amssymb,prb,version2,preprint,aps]{revtex}
\begin{document}
\tightenlines
\begin{title}
Central role of exchange-correlation hole \\
 in the 2D metal-insulator
transition
\end{title}
\author{J.S. Thakur and D. Neilson}
\begin{instit}
School of Physics, The University of New South Wales, Sydney 2052 Australia
\end{instit}
\begin{abstract}
We show that the metal to insulator transition, whether generated by
decreasing the electron density or by increasing the spin alignment, is
determined by a universal functional form of the two-electron
correlation function $g(r)$.  This result provides direct evidence of
the central role of the Coulomb repulsion and exchange in driving the
metal-insulator transition.

{73.20.Dx,71.30.+h,73.40.-c}
\end{abstract}

The discovery that electrons in a two-dimensional plane can switch from
insulating to metallic behaviour as their density is decreased
\cite{Kravchenko} was unexpected as it had been widely believed that
electrons should always remain insulating in 2D.
\cite{Kravchenko,Abrahams}  The recent observations that a magnetic
field applied parallel to the plane destabilizes the 2D metallic phase
at fixed density \cite{Simonian,Pudalov1,Hamilton,Hanein} is puzzling
since a parallel field only aligns the internal spins of the electrons
and does not affect their external orbital motion.  Since it is widely
believed that the electron interactions are a key to understanding the
transition it is important to understand the role of
exchange-correlations in these phenomena.  We show that the two
phenomena can be characterized within the single framework of a
universal form for the exchange-correlation hole at the transition.

At the relatively high densities found in conventional metals and
semiconductors the electron correlations resulting from mutual Coulomb
repulsion are not important because the average interaction energies
are much smaller than the Fermi energies.  Thus the transport
properties of conventional metals and semiconductors are well accounted
for by the standard {\em nearly-free-electron} picture.  However
without electron repulsion the 2D system would always be insulating in
the presence of disorder \cite{Abrahams} so it is clear that to treat
the metal-insulator transition we must look beyond this picture.

At low electron densities the Fermi energy is small and the
electron-electron correlations dominate.  For extremely low electron
densities the strong electron correlations drive the system to a
localized state (the Wigner crystal) \cite{Ceperley} without the need
for disorder.  The metal-insulator transition is observed at
significantly higher densities than the value predicted theoretically
in Ref.\ \onlinecite{Ceperley} for the Wigner transition.  This has
been interpreted as due to the effect of electron-defect interactions
since defects reduce the mobility of the electrons, making them easier
to localize.

In the Wigner crystal the strong correlations lead to near neutrality
within each Wigner-Seitz cell since each electron is surrounded by a
region of near-zero electron density. \cite{Ceperley}  This repulsive
hard core region forms part of the electron's exchange-correlation
hole.  The complete density profile of the exchange-correlation hole is
known from the two-electron correlation function $g(r)$ given in Ref.\
\onlinecite{Ceperley}.  When the electron density is increased, the
radius of the hard core in the exchange-correlation hole shrinks and
the Wigner crystal melts.  However the hard core persists in the
delocalized state until the density increases to $r_s\simeq7$, a factor
$25$ times greater than the Wigner transition density of
$r_s\simeq35$.  $r_s$ is the average electron spacing in effective Bohr
radii.

The diffusion quantum Monte Carlo numerical simulations (DQMC) in
Ref.\ \onlinecite{Ceperley} also show that the exchange-correlation
hole is stronger for spin polarized electrons than for unpolarized
electrons.  This is due to the additional exchange acting between the
increased number of parallel spin electrons.  In Fig. 1 we see at fixed
density that polarizing the spins significantly expands the size of the
hard core in $g(r)$.  At $r_s=7$ for the polarized system there is
still a region of zero density around each electron while for the
unpolarized system it has disappeared.

We proposes here that the suppression of the metallic state by a
parallel magnetic field is associated with the expansion of the hard
core due to alignment of the electron spins by the field.  We estimate
the degree of electron spin polarization as a function of a parallel
magnetic field $H_\parallel$ using DQMC data from Rapisarda and
Senatore \cite{Senatore}.  At these electron densities the free energy
$E_p(r_s)$ per electron for the fully spin polarized state is close to
the free energy $E_u(r_s)$ for the unpolarized state so that the Zeeman
energy gain from small magnetic fields is sufficient to fully polarize
the ground state.  We estimate the critical $H_\parallel$ needed to
fully polarize the electron spins by equating the Zeeman energy gain to
the difference in free energies,
$(g\mu_B/\hbar)H_c=[E_p(r_s)-E_u(r_s)]$.  With $(g\sigma_z)=1.1$ taken
for GaAs, \cite{Daneshvar} a field $H_\parallel\simeq 0.7$ T is
sufficient to produce full spin polarization for $r_s=9$.

In Hamilton {\it et al} 's \ \cite{Hamilton} recent experiment the
metal-insulator transition boundary in $p$-GaAs is shifted by a
parallel field $H_\parallel=0.6$ T from hole density
$p_s=7.5\times10^{10}$ cm$^{-2}$ to $p_s=12.4\times10^{10}$ cm$^{-2}$
that is from $r_s=9$ to $r_s=7$.  If we move along the experimental
transition line in the direction of decreasing $r_s$, the critical
magnetic field increases.  This results in the electron spins becoming
increasingly aligned.  From a linear interpolation between
$H_\parallel=0$ and $H_c$ we determine at $r_s=7$ that a field
$H_\parallel=0.6$ T induces 50\% spin polarization.  Figure 2 shows
that as a result of the increased polarization, the $g(r)$ on the
transition boundary at $r_s=7$ for the polarized system is essentially
identical with the $g(r)$ on the transition boundary at $r_s=9$ for the
unpolarized system.  The hard core radius remains fixed along the
entire experimental transition line.  This indicates that the
transition is determined by a unique functional form of the $g(r)$, or
equivalently, by a critical density profile of the exchange-correlation
hole.

Because of the enhancement in the exchange, at fixed $r_s$ the critical
impurity density $n_{i}$ for the polarized system is smaller than for
the unpolarized system .  In a separate calculation \cite{TN3} we have
shown that fully spin polarizing the system with an $H_\parallel\sim1$
T destabilizes the metallic phase.  

In conclusion, the free energy difference between the spin polarized
and unpolarized states for the electron densities of relevance here, is
so small that a magnetic field $\sim1$ T is sufficient to fully
polarize the system.  Using numerical simulation results we have shown
that the additional gain in exchange-correlation for the polarized
state is a key in understanding the instability of the 2D metallic
state.  Decreasing the electron density also enhances
exchange-correlation, which again destabilizes the metallic state.  We
show that the metal-insulator boundary is determined by a universal
functional form of the two-electron correlation function $g(r)$
independent of the degree of polarization and the electron density.

\acknowledgements

This work is supported by Australian Research Council Grant.  We thank
Mukunda Das, Michelle Simmons and Lessek \'{S}wierkowski for their
useful comments.

\figure{Two-electron correlation function $g(r)$ taken from Ref.\
\onlinecite{Ceperley} at $r_s=9$ for unpolarized system (dotted line)
and fully polarized system (dashed line), showing the effect of exchange
enhancement..
 \label{g1(r)}}

 \figure{Correlation function $g(r)$ along the metal-insulator
transition boundary.  Dotted line: $r_s=9$, unpolarized. Dash-dot line:
$r_s=8$ for $H_\parallel=0.4$ T.  Dashed line: $r_s=7$ for
$H_\parallel=0.6$ T.  The three curves are essentially identical.
 \label{g2(r)}}


\begin{references} 

\bibitem{Kravchenko} S.V. Kravchenko,  G.V. Kravchenko  and J.E.
Furneaux, \prb {\bf 50}, 8039 (1994); S.V. Kravchenko, Whitney E.
Mason, G.E.  Bowker, J.E. Furneaux, V.M. Pudalov and M. D'Iorio, \prb
{\bf 51} 7038 (1995); S.V. Kravchenko, D. Simonian, M.P. Sarachik,
Whitney E. Mason and J.E. Furneaux, \prl {\bf 77}, 4938 (1996);
Dragana Popovic, , A.B.  Fowler and S. Washburn, \prl {\bf 79}, 1543
(1997); M.Y. Simmons, A.R. Hamilton, M. Pepper, E.H. Linfield, P.D.
Rose, D.A. Ritchie, A.K.  Savchenko and T.G. Griffiths, \prl {\bf 80},
1292 (1998); M.Y.  Simmons, A.R. Hamilton, T.G. Griffiths, A.K.
Savchenko, M. Pepper and D.A. Ritchie, Physica B {\bf 251}, 705 (1998)

\bibitem{Abrahams}E. Abrahams, P.W. Anderson, D.C. Licciardello and
T.V. Ramakrishnan, \prl {\bf 42}, 673 (1979)

\bibitem{Simonian} D. Simonian, S.V. Kravchenko, M.P. Sarachik and
V.M. Pudalov, \prl {\bf 79}, 2304 (1997) 

\bibitem{Pudalov1} M. Pudalov, G. Brunthaler, A. Prinz and G. Bauer, 
Pis'ma Zh. Eksp. Teor. Fiz. {\bf 65}, 887 (1997) (JETP Lett. {\bf 65}, 932
(1997)) 

\bibitem{Hamilton} A.R. Hamilton, M.Y. Simmons, M. Pepper, E.H.,Linfield,
P.D. Rose and D.A. Ritchie,  \prl {\bf 82}, 1542 (1999)

\bibitem{Hanein}Y. Hanein, N. Nenadovic, D. Shahar, H. Shtrikman, 
J. Yoon, C.C. Li and D.C. Tsui,  Nature {\bf 400}, 735 (1999)

\bibitem{Ceperley} B. Tanatar and D.M. Ceperley, Phys. Rev. B {\bf 39},
5005 (1989) 

\bibitem{Senatore} Francesco Rapisarda and Gaetano Senatore, Aust J.
Phys. {\bf 49}, 161 (1996) 

\bibitem{Daneshvar} A.J. Daneshvar, C.J.B. Ford, M.Y. Simmons, 
A.V. Khaetskii, A.R. Hamilton, M. Pepper and D.A. Ritchie, \prl {\bf
79}, 4449 (1997) 

\bibitem{TN3} J.S. Thakur and D. Neilson, unpublished

 \end{references}
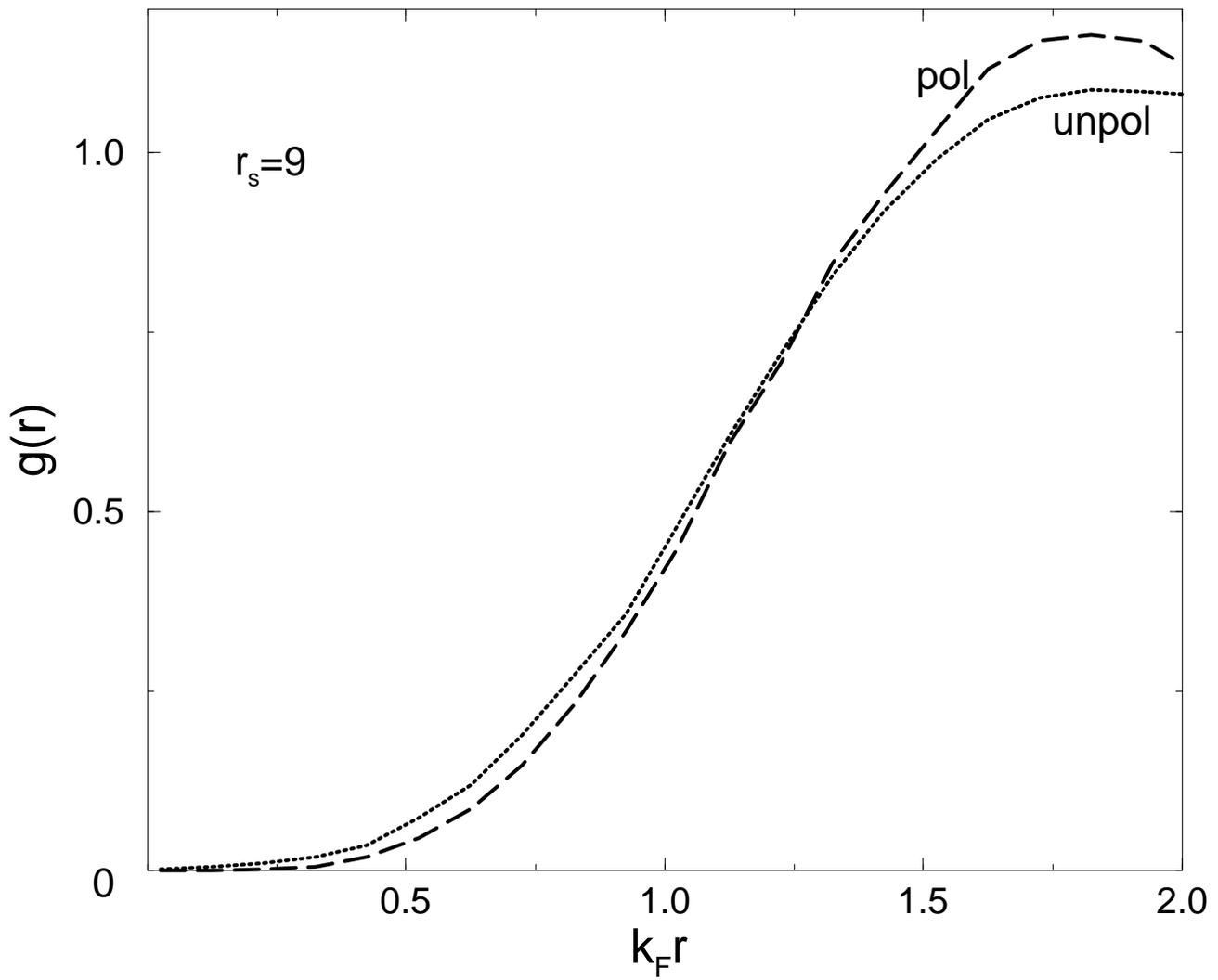
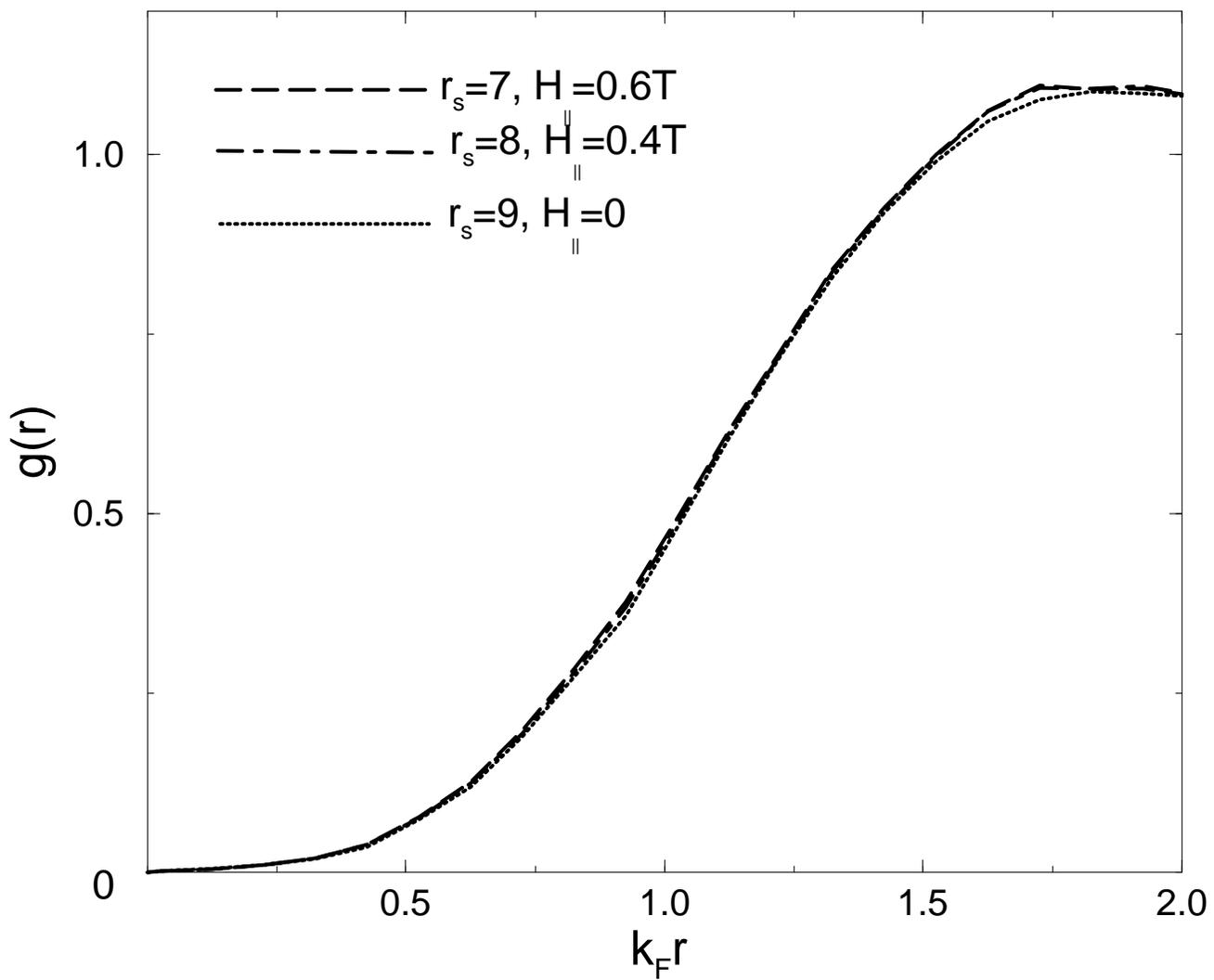
\end{document}